\documentclass{svproc}
\usepackage{cite}
\usepackage{graphicx}
\usepackage{geometry}
\usepackage{url}
\usepackage[font={small,bf}]{caption}
\usepackage[yyyymmdd]{datetime}

\begin{document}

%
\title{MusicTM-Dataset for Joint Representation Learning among Sheet Music, Lyrics, and Musical Audio}
\titlerunning{Hamiltonian Mechanics}  
%

%
%
\author{Donghuo Zeng \inst{1} \and Yi Yu\inst{2} \and 
Keizo Oyama\inst{3}}
\institute{$^{123}$National Institute of Informatics, SOKENDAI, Tokyo, Japan\\
\email{donghuozeng@gmail.com}}
\maketitle 

\begin{abstract}
This work present a music dataset named MusicTM-Dataset, which is utilized in improving the representation learning ability of different types of cross-modal retrieval (CMR). Little large music dataset including three modalities is available for learning representations for CMR. To collect a music dataset, we expand the original musical notation to synthesize audio and generated sheet-music image, and build musical notation based sheet-music image, audio clip and syllable-denotation text as fine-grained alignment, such that the MusicTM-Dataset can be exploited to receive shared representation for multi-modal data points. The MusicTM-Dataset presents 3 kinds of modalities, which consists of the image of sheet-music, the text of lyrics and synthesized audio, their representations are extracted by some advanced models. In this paper, we introduce the background of music dataset and express the process of our data collection. Based on our dataset, we achieve some basic methods for CMR tasks. The MusicTM-Dataset are accessible in~\url{https://github.com/dddzeng/MusicTM-Dataset}.
\end{abstract}

\section{Introduction}
Music data is getting readily accessible in digital form online, which brings difficult to manage the music from a large amount of personal collection. It highly relies on the music information retrieval to retrieve the right data information for users. In recent years, machine learning or deep learning based methods has become increasing prevailing in music information retrieval~\cite{eyben2010universal, hamel2010learning, zhou2015chord, bock2015accurate, grill2015music, choi2017tutorial, siedenburg2016comparison, sigtia2015audio} and has played an essential role in MIR.

\begin{figure}
    \centering
    \includegraphics[width=13.8cm, height=7cm]{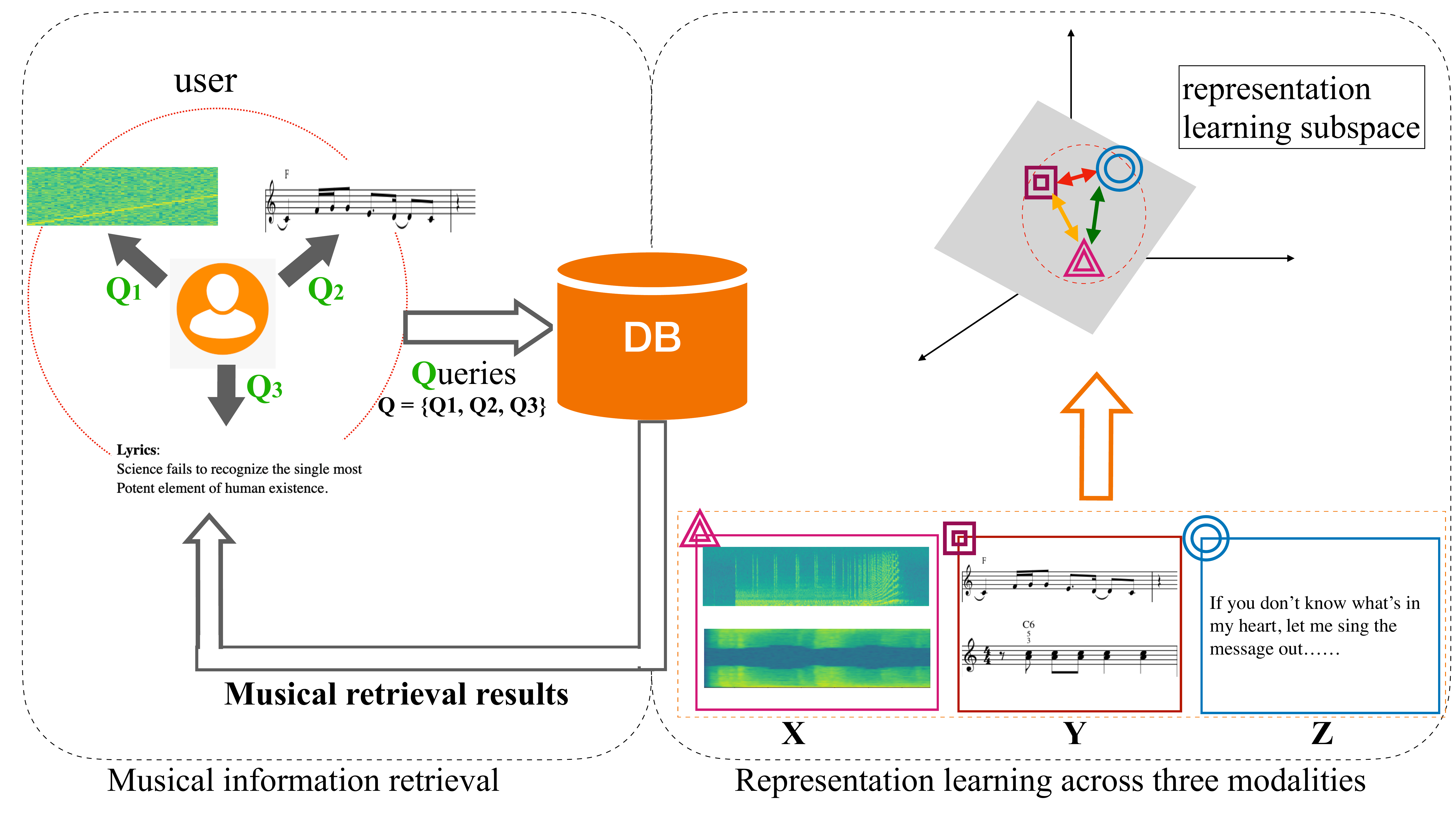}
    \caption{The framework of representation learning for music information retrieval.}
    \label{fig:mir}
\end{figure}

This paper concentrates on content music MIR by learning semantic concepts across different music modalities for MIR, as shown in the Fig.~\ref{fig:mir}. For instance, when we play music audio, we want to find what is the corresponding sheet music and which lyrics is correct, by learning two kinds of relationship in audio-sheet music and audio-lyrics. Such kinds of relationship obtained from content-based representation by learning the alignment across two modalities in the shared latent subspace without introducing any users' information. The unsupervised representation learning method ensures the system can allow users to find the right music data modalities with the other data modalities as query.

The major challenge of unsupervised representation learning for different music modalities is the modality gap. Representation learning for two music data modalities such as audio-lyrics~\cite{yu2019deep, kruspe2016retrieval, kruspe2018retrieval}, audio-sheet music~\cite{balke2019learning, dorfer2018learning}, have become increasingly in the CMR task to bridge the modality gap. In previous works, classic CCA and CCA-variant methods~\cite{dorfer2016towards, dorfer2018end, zeng2018audio, zeng2019learning, zeng2020deep} are popular in representation learning between two music data modalities, through finding linear or nonlinear transformation to optimize the correlation between two data modalities in the shared latent subspace. with the success of Deep Neural Network (DNN) in representation learning, DNN is also helpful for learning joint representation for cross-modal tasks~\cite{yu2018category}, for example, attention network~\cite{balke2019learning} applies a soft-attention mechanism for the audio branch to learn the relationship between sheet music and audio, which solves the problem that the music recordings easily brings about the global and local time deviations.

However, representation learning for two modalities is still not enough to achieve the music information retrieval, when we apply one data modality as query to retrieve other two different data modalities. The existing dataset normally applied in learning correlation between two modalities in a shared space. The paper~\cite{dorfer2018learning} collect a dataset contains an alignment between sheet music and music audio, which explores music audio to find the corresponding sheet music snippets. \cite{mauch2011integrating} apply a lyrics and audio paired dataset to align lyrics to audio. In this paper, we collect a new music dataset including three music data modalities. In particular, sheet music and audio are generated from music notes by music generation tools, the syllable-level lyrics and music notes are fine-grained alignment. Three major contributions of this paper have achieved in the following aspects: 1) we collect a fine-grained alignment across three music data modalities, which is useful for representation learning methods to obtain high-level feature for music CMR tasks. 2) we release experimental results of some baselines such as CCA and Generalized CCA on our MusicTM-Dataset. 3) The performance of Generalized CCA surpasses the CCA on audio-sheet music CMR task, which shows that the mapping all the three data modalities into a shared latent subspace can be better than mapping them into two shared latent subspace for audio-sheet music cross-modal retrieval.

The rest parts are arranged as follows. Some existing related works show in section 2. In section 3, we explain the detail of our data collection, feature representations and the metrics we applied on our experiment in section 4. Section 5 makes a conclusion of the whole paper.

\section{Related works}
\subsection{Audio and lyrics}
Recently, the study of automatic audio-lyrics alignment techniques is getting trendy. The aim of the topic is to estimate the relation between audio and lyrics, such as temporal relation~\cite{FujiharaG12}, deep sequential correlation~\cite{Yu08976}. \cite{mauch2011integrating} establishes audio-lyrics alignment based on a hidden Markov model speech recognizer, in particular, the lyrics input is to create a language model and apply the Viterbi method to link the audio and the lyrics. Synchronizing lyrics information with an audio
recording is an important music application. \cite{LeeS17} presents an approach for audio-lyric alignment by matching the vocal track and the synthesized speech.

\subsection{Sheet music and audio}
The popular problem of correlation learning between sheet music and audio is to establish the relevant linking structures between them. In ~\cite{ThomasFMC12}, it aims to establish linking the regions of sheet music to the corresponding piece in an audio of the same clip. ~\cite{DorferAW16} bring forwards an multi-modal convolutional neural network, by taking an audio snippet as input to find the relevant pixel area in sheet music image. However, the global and local tempo deviations in music recordings will influence the performance of the retrieval system in the temporal context. To address that, \cite{BalkeDCAW19} introduces an additional soft-attention mechanism on audio modality. Instead of correlation learning with high-level representations, \cite{dorfer2018learning} matches music audio to sheet music directly, the proposed method learns shared embedding space for short snippet of music audio and the corresponding piece in sheet music.

\subsection{Lyrics and sheet music}
Learning the correlation between lyrics and sheet music is a challenging research issue, which requires to learning latent relationship with high-level representations. The automatic composition techniques are considerable for upgrading music applications. \cite{yuHCMTJ20} proposed a novel deep generative model LSTM-GAN to learn the correlation in lyrics and melody for generation task. Similarly, \cite{fukayama2010automatic} presents an approach that is used to generate music song from a Japanese lyrics. \cite{WatanabeMFGIN18} introduces a novel language model that can generate lyrics from a given sheet music. \cite{WangJW10} presents an better query in using lyrics and melody, which take advantage of extra lyrics information by linking the scores from pitch-based lyrics and melody recognition. Accept that, “singing voice,” which is for generating singing voice has been drawing attention in the last years, \cite{liu2019score} explores a novel model that the singing voice generation with no consideration of pre-assigned melody and lyrics.

\section{Dataset and Metrics}
This section presents the motivation and contribution of our data collection. Moreover, also the process of dataset collection applied in our experiments and the data feature extraction are discussed. In the end, we show all the evaluation metrics applied to leverage our models.
\subsection{Dataset Collection}
\begin{figure*}[h]
   \centering
   \includegraphics[width=15.0cm, height=7.0cm]{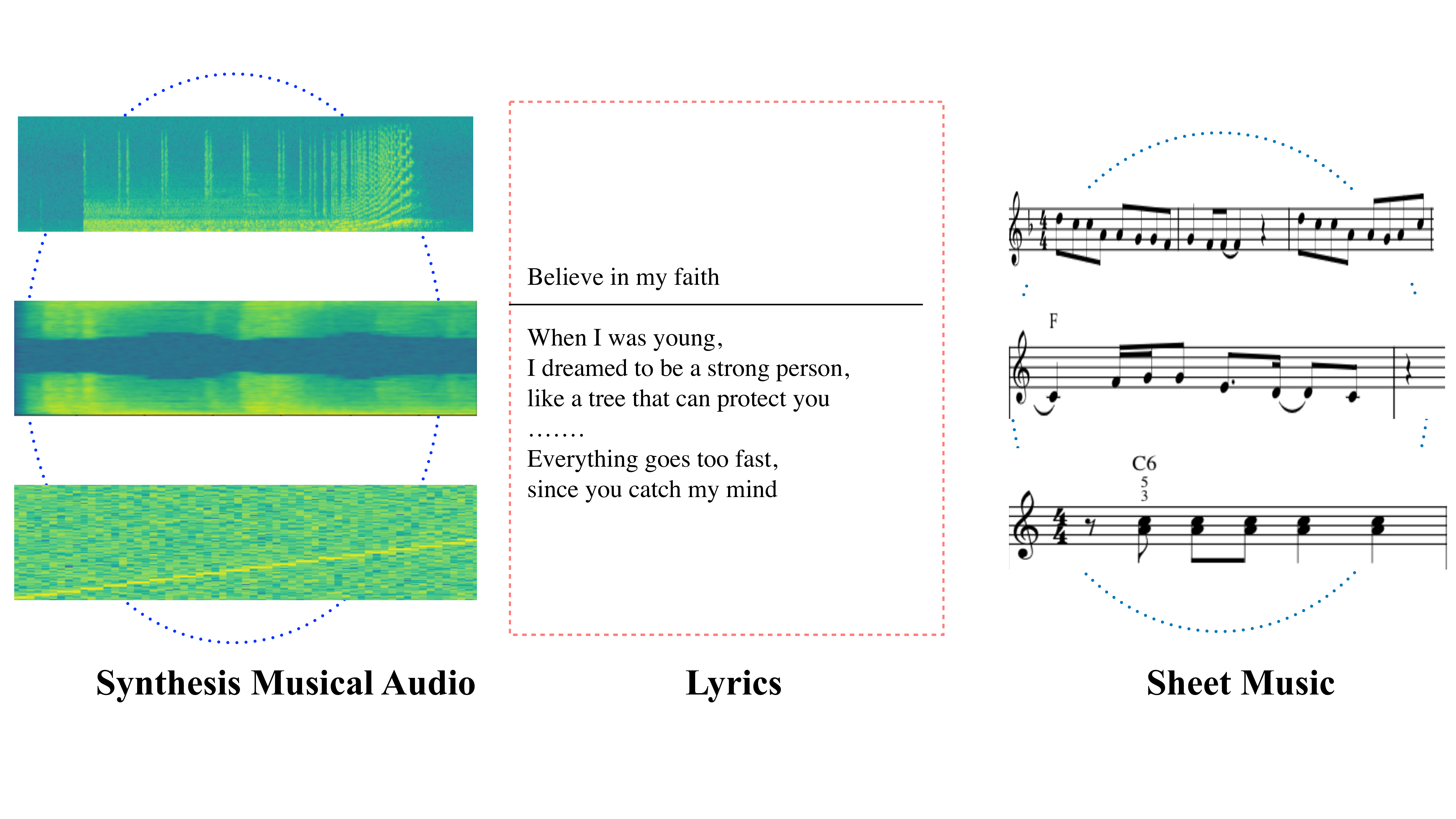}
   \caption{Examples of three data modalities in MusicTM-Dataset.}
   \label{fig:sample}
 \end{figure*}

Fig.~\ref{fig:sample} shows a few examples of MusicTM-Dataset we applied, including the spectrum of  music audio with Librosa library~\footnote{https://librosa.org/doc/latest/index.html}, word-level lyrics, and sheet music with Lilypond technique~\footnote{http://lilypond.org/}.

 \begin{figure*}[h]
   \centering
   \includegraphics[width=15.0cm, height=7.3cm]{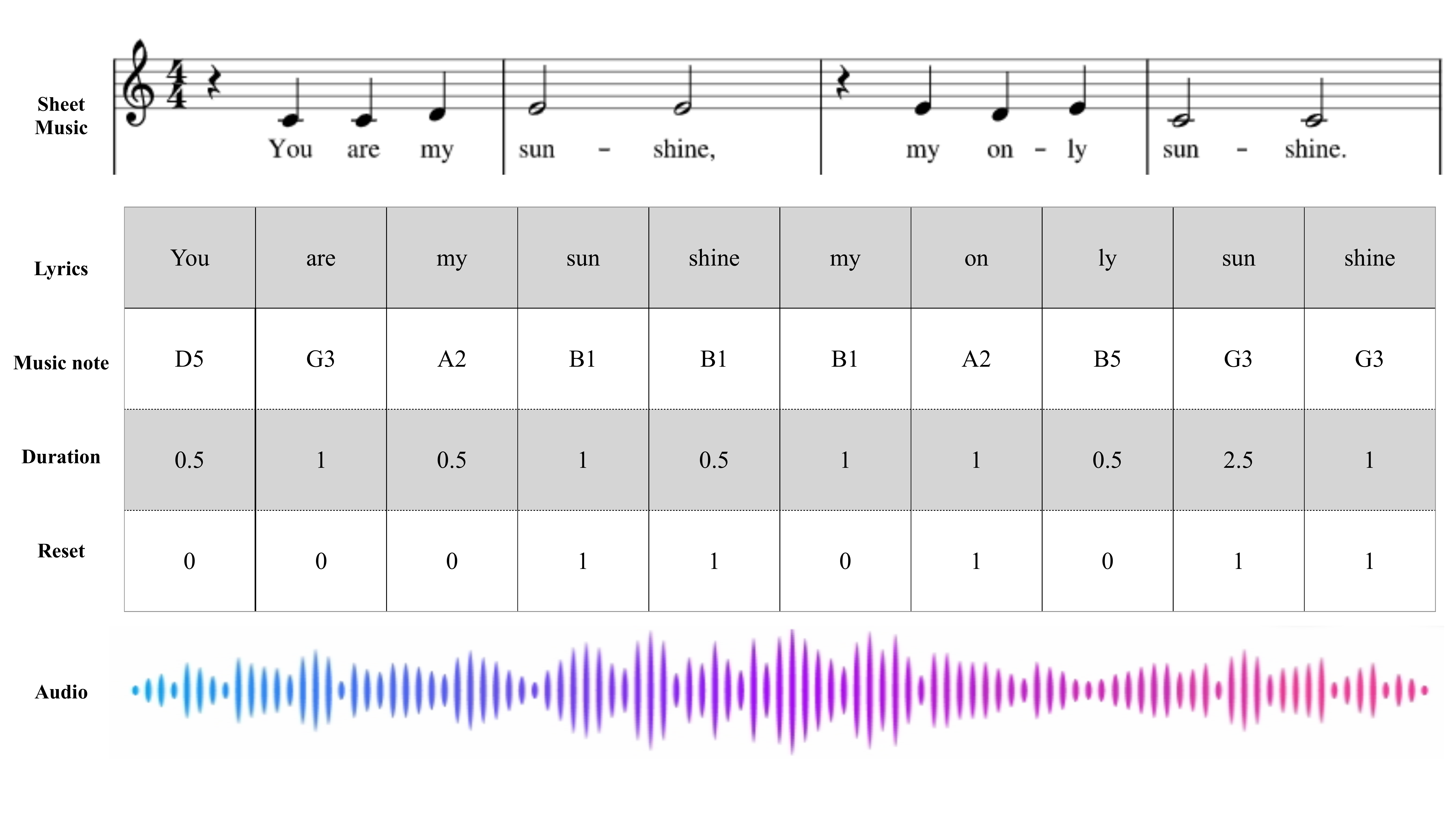}
   \caption{An example of fine-grained alignment across three modalities.}
   \label{fig:alignment}
 \end{figure*}
 
 The available music dataset with three modalities, which can be applied in music information retrieval based on the high-level semantic features is rarely reported. We try to learn aligned representation for sheet music images, music audio, and lyrics because they frequently appear in the music data collection. We follow the work~\cite{yuHCMTJ20} to collect our music dataset by extending two modalities (lyrics and music notes) to three modalities: sheet music, audio, and lyrics.

In~\cite{yuHCMTJ20} presents a music dataset that a music is represented by lyrics and music notes. the lyrics is parsed as syllable-level collection, such as the lyrics: 'Listen to the rhythm of fall ...' will parse as 'Lis ten to the rhy thm of fall'. A music note is a ternary structure that includes three attributions: pitch, duration, and rest. The pitch is a frequency-related scale of sounds, for example, piano keys MIDI number ranges from 21 to 108, each MIDI number corresponds to a pitch number, such as MIDI number '76' represents pitch number 'E5'. Duration in music notes denotes the time of the pitch, for example, a pitch number 'E5' with its duration 1.0, means this music note will last 0.5 seconds in the playing. The rest of the pitch is the intervals of silence between two adjacent music notes, which share the same unit with duration. The dataset used for the melody generation from lyrics, to consider the time-sequence information in the pairs, the syllable-level lyrics and music notes are aligned by pairing a syllable and a note. 

The initial pre-processing for our dataset is to get the beginning of music notes and corresponding syllables.  
In our MusicTM-Dataset collection, we adopted the same method to get the first 20 notes as a sample and ensure the syllable-level lyrics corresponding can be kept. Moreover, we removed the samples if existing the rest attributes of the note are longer than 8 (about four seconds). 

Music audio and sheet music are separately created from music notes that matches our purpose of musical multimodal building. We use syllable-level lyrics and notes to create the pairs of sheet and audio by some high-quality technologies. All the music data modalities contain temporal structure information, which motivates us to establish fine-grained alignment across different modalities, as seen in Fig.~\ref{fig:alignment}. In detail, the syllable of lyrics, the audio snippet, and sheet music fragment generated from music notes are aligned.

\textbf{Music audio} is also music sound transmitted in signal form. We add piano instrument in the music channel to create new midi files, and synthesize audios with TiMidity++ tool~\footnote{http://timidity.sourceforge.net/.}

\textbf{Sheet music} is created by music note with Lilypond tools. Lilypond is a compiled system that runs on a text file describing the music. The text file may contain music notes and lyrics. The output of Lilypond is sheet music which can be viewed as an image. Lilypond is like a programming language system, music notes are encoded with letters and numbers, and commands are entered with backslashes. It can combine melody with lyrics by adding the "$\backslash$addlyrics" command. In our MusicTM-Dataset, sheet music (visual format) for one note and entire sheet music (visual format) for 20 notes are created respectively. Accordingly, each song has single note-level and sequential note-level (sheet fragment) visual formats.

\subsection{Feature Extraction}
This section will explain the feature extraction for music multimodal data.
\subsubsection{Audio Feature Extraction}
\begin{figure*}[h]
  \centering
  \includegraphics[width=15.0cm, height=2.1cm]{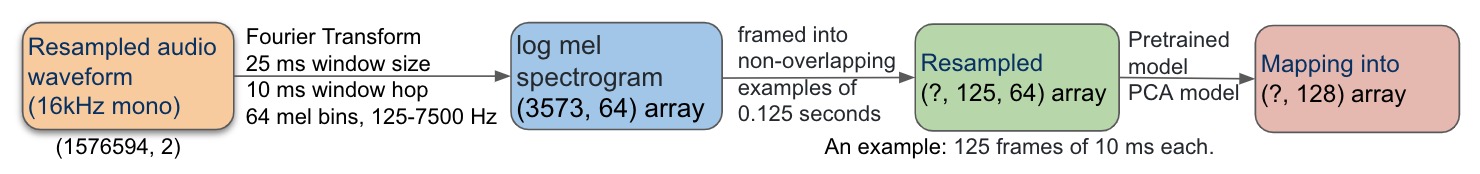}
  \caption{The audio feature extraction process with vggish model}
  \label{fig:audio}
\end{figure*}

Generally, audio signal is used for audio feature extraction, which plays the main role in speech processing~\cite{watanabe1804espnet, lambert2017task}, music genre classification~\cite{kobayashi2018audio}, and so on. Here, we present a typical model for audio feature extraction, the supervised trained model Vggish. The detailed process of feature extraction can be seen in Fig.~\ref{fig:audio}. Firstly, we resample audio waveform 16 kHz mono, then calculate a spectrogram. 
Secondly, in order to obtain a stable log mel spectrogram, it is computed by exploring log. Finally, resampling the feature into (125,64) format, then applying pre-trained model to extract feature and use PCA model to map it into 128-dimensional.

\subsubsection{Sheet music Feature Extraction}
Different from other image feature extraction, our feature extraction of sheet music image tries to catch pitches and the segments. In this paper, our information extraction of sheet music has two levels, pitch detection, and semantic segments. We apply the ASMCMR~\cite{DorferAW17} model trained in audio-sheet retrieval tasks, which learns the correlation between audio clips and corresponding sheet snippets. In our work, the shape of extracted note-level feature and sheet snippet-level features are (100, 32) and (32,) respectively.

\subsubsection{Lyrics Feature Extraction}
We follow~\cite{yuHCMTJ20} to keep the alignment between syllable and note by representing lyrics in the form of syllable and word level. The syllable-level feature extracted with the syllable skip-gram model, the word-level feature extracted with the word skip-gram model used in~\cite{yuHCMTJ20}. These two pre-trained skip-gram models are trained on all the lyrics data, which applied in a regression task with SGD optimization. The input of syllable-level skip-gram model is a sequence of syllables in a sentence, while the input of word-level model is a word unit sequence in the sentence. The output of the syllable-level and word-level skip-gram model is 20-dimensional embedding for each syllable and word, respectively.

The overall statistics of our music data are shown in Table~\ref{tab:musicdataset}. We divided the dataset into 3 parts as training, validation, and testing set by 70\%, 15\%, and 15\%. The number of training, validation, and testing set are 13,535, 2800, and 2800 respectively.

\begin{table}
\begin{center}
  \caption{Statistics of MusicTM-Dataset applied in our experiments}
  \label{tab:musicdataset}
  \begin{tabular}{c|ccl}
    Modality &Feature Extractor &Dimension &Number\\
    \hline
    Audio & Vggish &(20, 128) &14,454\\
    Lyrics & Skip-gram &(20, 20) &14,454\\
    Sheet music & Lilypond\&ASMCMR & (20, 100, 32) &14,454\\
\end{tabular}
\end{center}
\end{table}

\subsection{Evaluation Metric}
To evaluate some baselines on our dataset, we apply some standard evaluation from the work~\cite{OtaniNRHY16} for unsupervised learning based cross-modal retrieval. 
R@K (Recall at K, here we set K as 1,5, and 10) is to compute correct rate that is the percentage of retrieved items corresponding to the query in the top-K of rank list. Fox instance, R@1 calculate the percentage of sample appear in the first item of retrieved list. In order to further evaluate our collected dataset with some baselines, we also apply the Median Rank and Mean Rank to compute the mean and median rank of all the correct results.

\section{Experiments}
\begin{table}
\begin{center}
  \caption{The Performance of Multimodal Information Retrieval on MusicTM-Dataset.}
  \label{tab:cross-modal}
  \begin{tabular}{c|ccccl}
    \hline
    \multicolumn{6}{c}{audio2lyrics retrieval}\\\hline
    Methods &R@1 &R@5 &R@10 &MedR &MeanR\\
    \hline
     Random Rank~\cite{zeng2020unsupervised} &0.028 &0.055 &0.076 &7312.0 &7257.2 \\
     CCA~\cite{HardoonSS04} &0.306 &0.350 &0.353 &423.0 &639.4\\
     GCCA~\cite{tenenhaus2011regularized} &0.040 &0.074 &0.093 &770.0 &881.1\\
    \hline
    \multicolumn{6}{c}{lyrics2audio retrieval}\\
    \hline
     Random Rank &0.027 &0.055 &0.076 &7316.0 &7257.3 \\
     CCA &0.304 &0.349 &0.354 &427.0 &639.3\\
     GCCA &0.039 &0.078 &0.095 &774.0 &881.6 \\
  \hline
  \multicolumn{6}{c}{sheet music2lyrics retrieval}\\ 
    \hline
     Random Rank &0.027 &0.055 &0.075 &7311.0 &7257.3 \\
     CCA &0.093 &0.172 &0.203 &524.0 &708.7\\
     GCCA &0.089 &0.0142 &0.167 &573.0 &770.5 \\
    \hline
    \multicolumn{6}{c}{lyrics2sheet music retrieval}\\
    \hline
    Random Rank &0.027 &0.055 &0.077 &7313.0 &7257.4 \\
     CCA &0.093 &0.168 &0.198 &522.0 &709.0\\
     GCCA &0.098 &0.014 &0.168 &578.0 &769.8\\
  \hline
  \multicolumn{6}{c}{audio2sheet music retrieval}\\ 
    \hline
     Random Rank &0.028 &5.57 &7.50 &7310.0 &7257.2 \\
     CCA &0.303 &0.349 &0.353 &341.0 &596.5\\
     GCCA &0.358 &0.403 &0.414 &271.0 &382.8 \\
    \hline
    \multicolumn{6}{c}{sheet music2audio retrieval}\\
    \hline
     Random Rank &0.026 &0.055 &0.075 &7310.0 &7257.4 \\
     CCA &0.300 &0.350 &0.354 &332.0 &596.1\\
     GCCA &0.362 &0.407 &0.415 &271.0 &381.3 \\
  \hline
\end{tabular}
\end{center}
\end{table}
\subsection{Baselines}
\textbf{CCA} can be seen as the method that aims at finding
linear transforms for two sets of variables in order to optimize the relation between the projections of the variable sets into a shared latent subspace. Consider two variables from two data modalities $X\in R^{D_{x}}$ and $Y\in R^{D_{y}}$ with zero mean and the two paired data sets $S_{x} = \{x_{1}, x_{2}, ..., x_{n}\}$ and $S_{y} = \{y_{1}, y_{2}, ..., y_{n}\}$. $W_{x}\in R^{D_{x}}$ and $W_{y}\in R^{D_{y}}$ as the directions that linearly map the two set into a shared latent subspace, such that the relation between the projection of $S_{x}$ and $S_{y}$ on $W_{x}$ and $W_{y}$ is optimized.

\begin{eqnarray}
\rho  = \arg\max_{(W_{x}, W_{y})} \frac{W^{T}_{x}\Sigma_{xy}W_{y}}{\sqrt{W^{T}_{x}\Sigma_{xx}W_{x}\cdot W^{T}_{y}\Sigma_{yy}W_{y}}} 
\end{eqnarray}

where $\rho$ is the correlation, $\Sigma_{xx}$ and $\Sigma_{yy}$ denote the variance–covariance matrix of $S_{x}$, $S_{y}$, respectively and $\Sigma_{xy}$ represents the cross-covariance matrix.

\textbf{Generalized CCA}~\cite{tenenhaus2011regularized} can be viewed as an extension method of CCA, which aims to solve the limitation on the number of data modalities. The objective function in Eq.~\ref{eq:Gcca}, which focuses on finding a shared representation $G$ for $K$ different data modalities.
\begin{eqnarray}
minimize_{(W_{k}, G)}  = \sum_{k=1}^{K} ||G - W_{k}^{T} X_{k}||_{F}^{2}
\label{eq:Gcca}
\end{eqnarray}
where $K$ is the size of data points, and $X_{k}$ is a matrix for $k^{th}$ data modality. Similar to CCA, GCCA is to find linear transformation for different data modalities to optimize the correlation within them.

\subsection{Results}
In Table~\ref{tab:cross-modal}, when learning the correlation between two data modalities with CCA method, the correlation of audio-lyrics and audio-sheet music can get more than 30\% of R@1, which illustrates the dataset can be learned for cross-modal retrieval task. Specifically, in comparison with CCA and RANDOM, GCCA will have a big improvement in the performance of audio-sheet music cross-modal retrieval. In detail, compared with CCA method, 5.46\% , 5.39\%, 6.06\%, 70, and 213.68 improved in R@1, R@5, R@10, MedR, and MeanR for music audio as the query to retrieve the correct sheet music; 6.16\%, 5.65\%, 6.1\%, 61, and 214.8 improved in R@1, R@5, R@10, MedR, and MeanR for sheet music as the query to retrieve the correct  music audio. However, GCCA will decrease the performance of audio-lyrics cross-modal retrieval and achieve a similar performance of sheet music-lyrics cross-modal retrieval.

The results show that the learned representation with GCCA for sheet image, lyrics, and music audio can raise the relation of sheet music and music audio. However, such representations drop the correlation between music audio and lyrics and their correlation between sheet music image and lyrics will almost stay the same as CCA method, which learns the representation in the shared subspace without involving lyrics data. The results prove our hypothesis can be accepted that the sheet music and music audio are created by music notes, so the correlation between audio and sheet music will be close. The lyrics and music note from original dataset exist alignment between each other, the correlation between the two can be learned. In this case, the correlation between audio and lyrics reflects the correlation between audio and music note, however, the correlation between sheet music and lyrics seems hard to learn. 

\begin{figure*}[h]
  \centering
  \includegraphics[width=15.0cm, height=7.5cm]{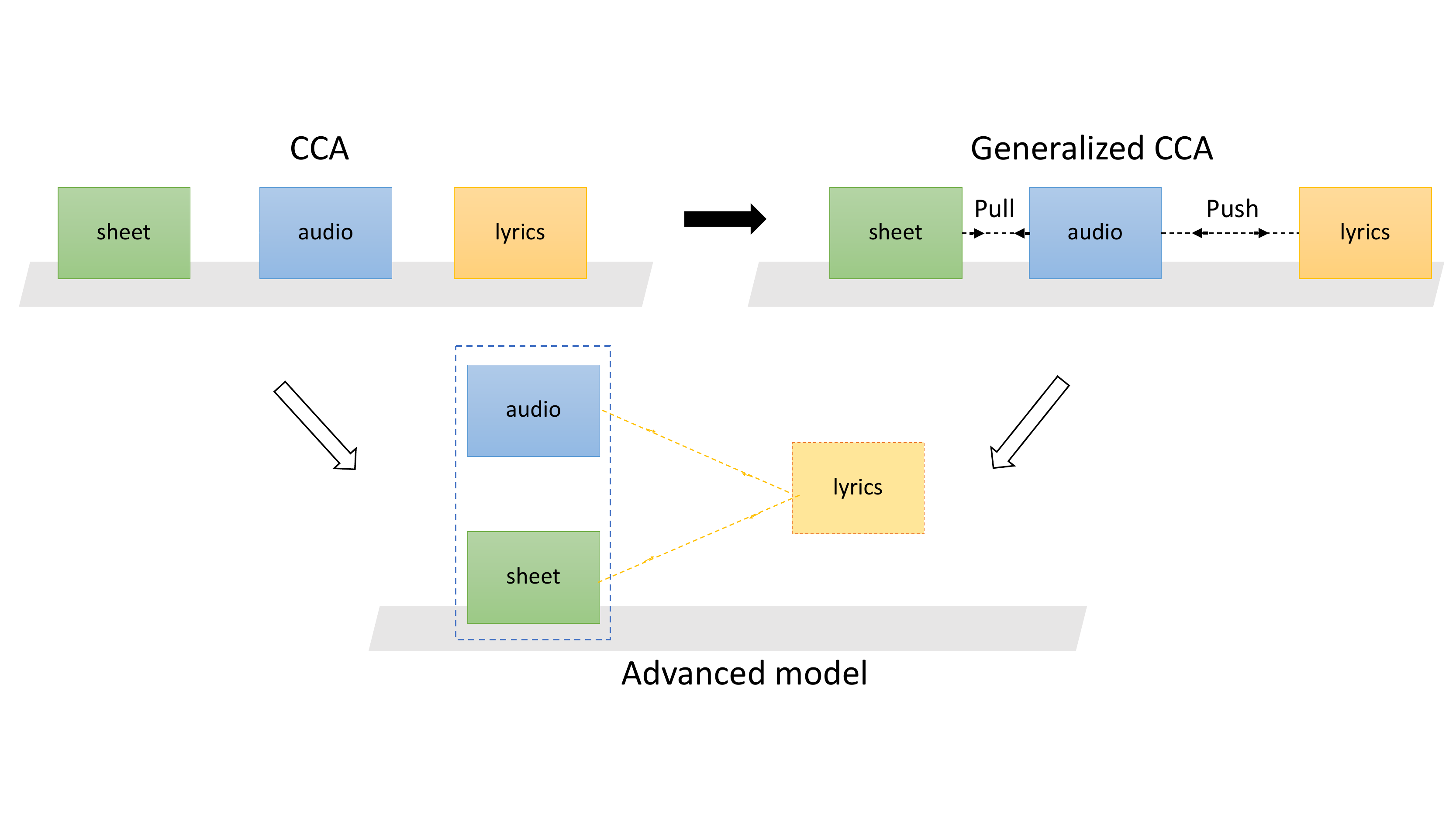}
  \caption{The general paradigm of MusicTM-Dataset with two different models (CCA, GCCA)}
  \label{fig:cca_gcca}
\end{figure*}

In visualization of the the position of sheet music, lyrics, and music audio in CCA and GCCA subspace, as shown in Fig.~\ref{fig:cca_gcca}. GCCA seems to pull audio and sheet music while pushing the audio and lyrics compared with the CCA subspace. This motivates us to propose a new advanced model that can improve three couples of cross-modal retrieval tasks in a shared latent subspace as the GCCA subspace achievement in the future. 

\section{Conclusion}
This paper presents a MusicTM-Dataset that consists of three different data modalities and there is fine-grained alignment across the modalities. The dataset can be easily extended to different researches, we report the performance of some baselines on our MusicTM-Dataset, which allows the results of the following research to be compared. Instead of applying CCA to learn shared latent subspace for every two modalities, GCCA learns the correlation of three modalities in one shared latent subspace. The performance of audio-sheet music can be improved and the performance of audio-lyrics cross-modal retrieval is quilt similar but the performance of lyrics-sheet music cross-modal retrieval will be decreased. In theory, we want to develop a new architecture that will improve the performance of multimodal information retrieval across different modalities.
\section*{Acknowledgements}
The JSPS Grant for SR financed this work, which is under Grant No. 19K11987.

\bibliographystyle{unsrt}
\bibliography{mybib}

\end{document}